\documentclass[11pt]{amsart}
\usepackage{geometry}                % See geometry.pdf to learn the layout options. There are lots.
\usepackage{graphicx}
\usepackage{amssymb}
\usepackage{epstopdf}
\usepackage{hyperref}
\pdfoutput=1
\title{Sticking around: an up-close look at droplet adhesion}
\author{Adam Paxson, Kripa K. Varanasi \\
Massachusetts Institute of Technology, Cambridge}
\begin{document}
\pagestyle{empty}
\maketitle
\begin{abstract}
We present a fluid dynamics video showing the adhesion of a drop to a superhydrophobic surface. We use environmental scanning electron microscopy to observe depinning events at the microscale. As the drop moves along the surface, the advancing portion of the contact line simply lies down onto the upcoming roughness features, contributing negligibly to adhesion. After measuring the local receding contact angle of capillary bridges formed on a micropillar array, we find that these depinning events follow the Gibbs depinning criterion. We further extend this technique to two-scale hierarchical structures to reveal a self-similar depinning mechanism in which the adhesion of the entire drop depends only on the pinning at the very smallest level of roughness hierarchy. With this self-similar depinning mechanism we develop a model to predict the adhesion of drops to superhydrophobic surfaces that explains both the low adhesion on sparsely structured surfaces and the surprisingly high adhesion on surfaces whose features are densely spaced or tortuously shaped. 
\end{abstract}
\section{Brief Explanation of Video Submission}
%\subsection{}
This video depicts the imaging technique used to visualize droplet pinning and depinning on discrete textured superhydrophobic surfaces, and some examples of motion of drops across different superhydrophobic textures. A Zeiss EVO-55 ESEM was used for imaging. A custom fixture was fabricated to mount the samples vertically in the ESEM and provide cooling with a Deben Coolstage Mk II Peltier cooling stage. A 10 ml water drop was affixed to the end of a copper wire that allowed the drop to be held against a vertically oriented surface, and also provided a cooling flux to minimize evaporation. By rotating the motorized stage of the ESEM, the droplet was swept across the surface at a velocity of 2 mm/s while observing the depinning events of the contact line at micron length scales. The chamber of the ESEM was maintained at a pressure of 1,000 Pa.

\section{References}
A. T. Paxson, K. K. Varanasi, "Self-similarity of contact line depinning from textured surfaces," Nature Communications, 4, 1492, 2013.
\end{document}